# Developing a single-phase and nanograined refractory high-entropy alloy ZrHfNbTaW with ultrahigh hardness by phase transformation via high-pressure torsion

Shivam Dangwal[1,2] and Kaveh Edalati[1,2]*

[1] WPI, International Institute for Carbon-Neutral Energy Research (WPI-I2CNER), Kyushu University, Fukuoka 819-0395, Japan
[2] Department of Automotive Science, Graduate School of Integrated Frontier Sciences, Kyushu University, Fukuoka 819-0395, Japan

High-entropy alloys (HEAs) are potential candidates for applications as refractory materials. While dual-phase refractory HEAs containing an ordered phase exhibit high hardness, there is high interest in developing intermetallic-free and single-phase refractory HEAs with high hardness. In this study, a new equiatomic HEA ZrHfNbTaW with an ultrahigh hardness of 860 Hv is developed. The alloy is first synthesized with a dual-phase structure via arc melting and further homogenized to a single body-centered cubic (BCC) structure by phase transformation via high-pressure torsion (HPT), using the concept of ultra-severe plastic deformation process. The ultrahigh hardness of the alloy, which is higher than those reported for refractory alloys and single-phase HEAs, is attributed to (i) solution hardening by severe lattice distortion, (ii) Hall-Petch grain boundary hardening by the formation of nanograins with 12 nm average size, and (iii) dislocation hardening confirmed by high-resolution transmission electron microscopy.

*Corresponding author (E-mail: kaveh.edalati@kyudai.jp; Tel/Fax: +81 92 802 6735)

**Introduction**

Since ancient times, engineering materials have been traditionally fabricated from one element (e.g. copper, gold) or two principal elements (e.g. bronze) together with adding minor amounts of other elements to enhance their properties (e.g. adding carbon to iron). A class of alloys known as high-entropy alloys (HEAs) was introduced in 2004, where multiple principal elements (at least five) are mixed to form an alloy [1,2]. By definition, HEAs are typically alloys with a high configurational entropy of more than 1.5$R$ ($R$: gas constant) [3,4]. The high-entropy effect reduces the Gibbs energy which enables the formation of a stable solid-solution phase, especially at high temperatures. The introduction of HEAs and their four core principles (high-entropy effect, sluggish diffusion, lattice distortion and cocktail effect) opened a path for the discovery of new materials with promising mechanical and functional properties [2,5]. HEAs have been investigated in different fields such as hydrogen storage [4,6,7], superconductors [8,9], nuclear applications [10] and high-temperature applications [11].

The alloys for high-temperature applications, which are known as refractory materials, should have the properties of high melting temperature, high hardness, high wear resistance and high corrosion resistance [12]. Ordered intermetallic compounds show good properties as refractory materials, but their main drawback is limited plasticity. Moreover, the ordered refractory materials sometimes form new phases at high temperatures, causing deterioration in mechanical, thermal and corrosion resistance properties [13,14]. The entropy-stabilization effect in HEAs and medium-entropy alloys (MEAs) can overcome the stability issue at high temperatures while producing high hardness combined with ductility. These features of HEAs have led to the introduction of refractory HEAs and MEAs in recent years [15,16].

Various refractory HEAs have been introduced such as TaNbHfZrTi [17], NbTaTiVW [18], VNbMoTaW [15] and MoNbTaTiW [19] which exhibit better ductility than intermetallics [20]. These HEAs have usually body-centered cubic (BCC) structures, as this phase offers promising potential in high-temperature environments [21–24]. Moreover, HEAs with the BCC structure are typically harder than HEAs with a face-centered cubic (FCC) structure which makes them more wear-resistant [15,17,18,25,26]. The hardness of HEAs can be further enhanced by grain refinement using severe plastic deformation (SPD) techniques [27,28] and there have been attempts to apply SPD to AlFeMgTiZn [29], CoCrFeMnNi [30–32], TiZrNbHfTa [33], etc. The hardening of HEAs by SPD is mainly due to grain refinement to the nanometer level and the contribution of the famous Hall-Petch equation [34,35,36]. While breaks in the Hall-Petch relationship or inverse Hall-Petch effect are observed in SPD-processed pure metals with nanograin sizes [37], such phenomenon is not expected to occur in refractory HEAs due to their high melting points.

There are various SPD methods such as accumulative roll-bonding [38], equal-channel angular pressing [39], high-pressure torsion (HPT) [40] and twist extrusion [41]. The HPT method, which was first introduced by Bridgman in 1935 [40], has been the most popular SPD method for processing HEAs. For the details of SPD methods, the readers are referred to a recent review paper authored by 122 experts in the field [42]. In the HPT process, torsional shear strain under very high pressures in the giga-Pascal range is applied on a cylindrical disc as shown in Fig. 1(a) [43,44]. The high pressure allows the processing of hard materials such as refractory HEAs, while a large shear strain can lead to grain refinement, chemical homogenization and phase transformations even in ultrahard and brittle materials [45]. The level of homogeneity and phase transformation in HPT processing is closely related to the degree of shear strain which is calculated by the following relationship [44].

$$\gamma = \frac{2\pi r N}{h} \tag{1}$$

where $\gamma$ is the shear strain, $r$ is the radial distance from the center of the disc rotation, $N$ is the number of turns, and $h$ is the height of the HPT-processed disc [44]. Using HPT on HEAs has resulted in notable enhancement in hardness, with measurements reaching up to 1000 Hv [46-48]. Moreover, HPT with extremely large shear strains, known as ultra-SPD [49,50], has been successfully used to synthesize various single-phase nanomaterials including HEAs [51,52].

In this study, the potential of HPT in grain refinement [53-56], phase transformation [57,58] and synthesis of single-phase alloys [49-52] is employed to produce a new refractory HEA with an equiatomic composition of ZrHfNbTaW, a single BCC phase, a nanograined structure with an average size of 12 nm and an ultrahigh hardness of 860 Hv. The successful application of HPT in this study introduces this method as a new path to explore refractory HEAs.

**Experimental Procedures**

Ingot of ZrHfNbTaW with equal atomic fraction of elements and 12 g mass was fabricated by arc melting of high-purity zirconium (99.2%), hafnium (99.7%), niobium (99.9%), tantalum (99.9%) and tungsten (99.9%). The synthesis was done in a water-cooled copper mold under an argon atmosphere. The ingot was rotated and re-melted seven times to ensure compositional homogeneity. For HPT processing, the ingot was cut into the form of cylindrical discs with 10 mm diameter and 0.8 mm height using a wire-cut electrical discharge machine. HPT processing was done at room temperature ($T$ = 300K) under a hydrostatic pressure of $P$ = 6 GPa and an angular speed of $\omega$ = 1 rpm. The torsional shear strain during the process was controlled by changing the number of turns ($N$) to 1, 10 and 50. The shear strain introduced in the HPT-processed disc after 50 turns is shown in Fig. 1(b) using a color contour. The temperature rise during HPT processing was measured using a thermocouple located inside the upper anvil of the HPT machine at 10 mm away from the disc center. The temperature rise in the sample was below 50 K which is negligible to influence the microstructure or phase transformation during the process, an issue that is consistent with earlier publications [47,59,60]. Following the HPT process, the alloy was examined by different methods, as described below.

First, an online melting temperature predictor software was used to estimate the melting point of the current alloy as well as of other refractory HEAs reported in the literature [61].

Second, the initial ingot ($N$ = 0) and HPT-processed discs ($N$ = 1, 10 and 50) were ground using emery papers (grit: 400, 800, 1200 and 2000). The crystal structure of the four samples was examined by X-ray diffraction (XRD) using Cu K$\alpha$ radiation with an acceleration voltage of 45 kV and a filament current of 200 mA. To determine the lattice parameters, the Rietveld method was used in the PDXL software.

Third, to measure the hardness of the samples, the discs were polished using a buff with alumina slurry to achieve a mirror-like surface on both sides of the disc. Vickers hardness testing machine was used to measure the hardness using a load of 0.5 kgf. The hardness in four radial directions was measured in positions spaced 1 mm apart as shown in Fig. 1(b). The average of these four measurements was taken into account as the average hardness in each radial distance from the disc center.

Fourth, the thermal stability of the HEA was examined by differential scanning calorimetry (DSC) analysis. A 3 mm disc with about 30 mg mass was cut from the HPT sample and processed for $N$ = 50. The sample was subjected to DSC with a heating rate of 5 K/min for up to 873 K. The sample after DCS was additionally examined by XRD.

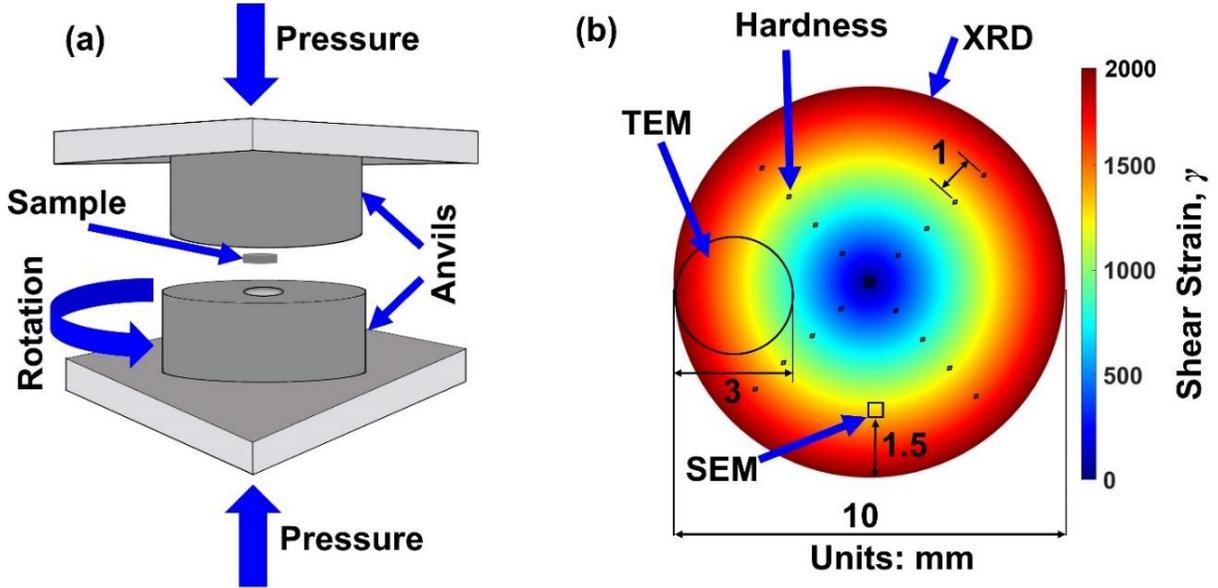

Figure 1. (a) Illustration showing high-pressure torsion (HPT) schematic: sample is a cylindrical disc, upper anvil is fixed whereas the lower anvil moves upward to apply high pressure and rotates to apply shear strain. (b) Color contour of the shear strain distribution in HPT-processed disc after $N = 50$ rotations.

Fifth, the microstructural and compositional analysis was conducted using field emission scanning electron microscopy (SEM) with an acceleration voltage of 15 keV. The discs were ground with emery papers (grit: 400, 800, 1200, 2000), subsequently polished with 9 µm and 3 µm diamond suspensions, and finally fine-polished by colloidal silica with 60 nm particle size to achieve a mirror-like polished surface. Energy dispersive X-ray spectroscopy (EDS) was done for compositional analysis and electron backscatter diffraction (EBSD) was used to determine the grain size of the initial ingot. The analysis of EBSD data was done using the MTEX toolbox in MATLAB [62] and grains smaller than 3 pixels were merged into the surrounding larger grains. The smoothing of grain boundaries was done by assigning the average orientation of surrounding grains to un-indexed grains using grain object property meanOrientation [62]. To measure the average grain size, both low-angle grain boundaries with misorientation angles between 2º and 15º and high-angle grain boundaries with misorientation angles more than 15º were considered.

Sixth, the nanostructure of the sample processed for $N = 50$ was examined by transmission electron microscope (TEM) and scanning-transmission electron microscopy (STEM). A disc of 3 mm diameter was cut at 2-5 mm from the center of the HPT-processed disc. A solution of 65 % methanol, 30 % butanol and 5% perchloric acid was used as an electrolytic solution for electrochemical polishing at 263 K under an applied voltage of 15 V. After electrochemical polishing, ion milling was done for 20 minutes using argon ions with an energy of 5 eV with incident angles of ±5º. High-resolution imaging, fast Fourier transform (FFT), high-angle annular dark-field (HAADF) imaging and EDS were used to analyze the sample.

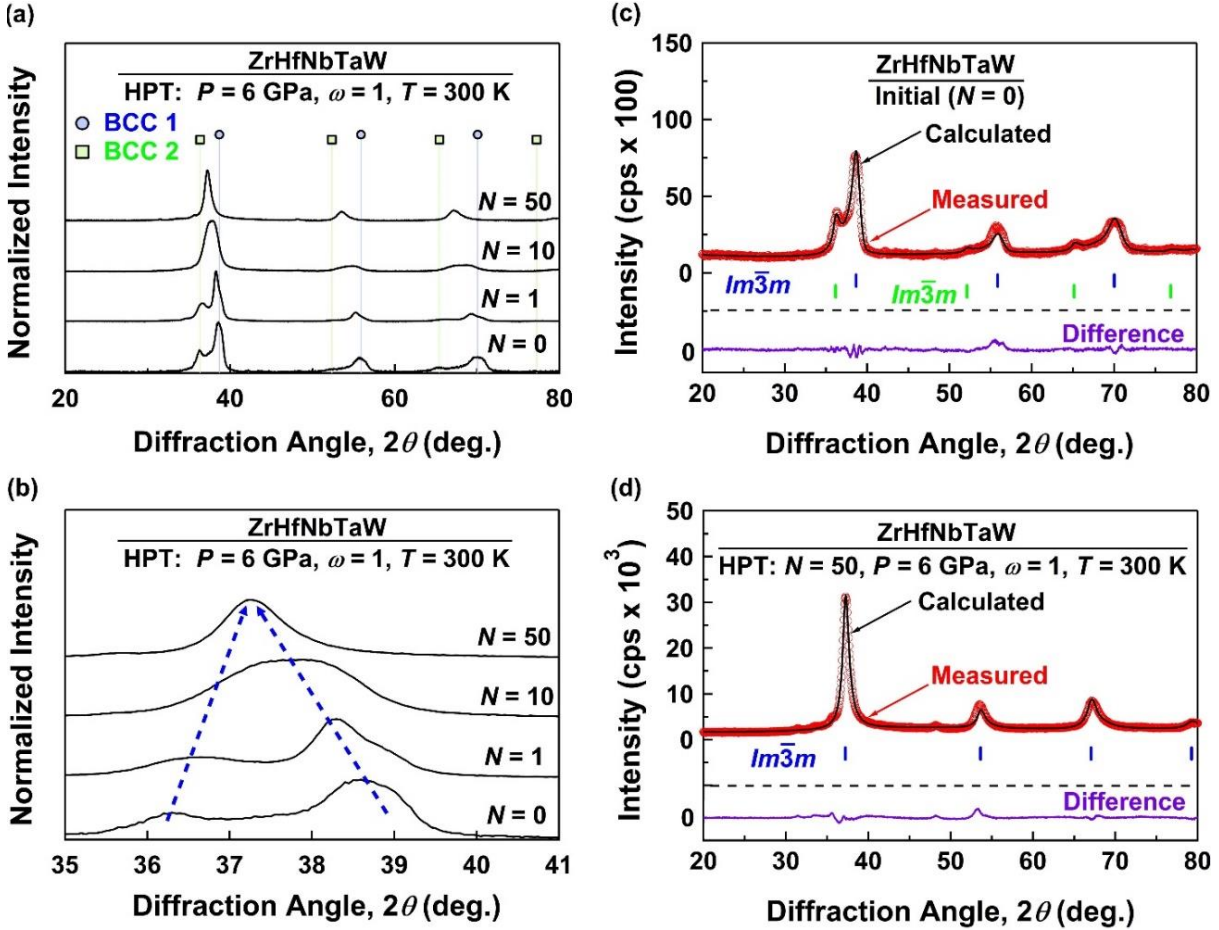

Figure 2. Strain-induced phase transformation from two BCC structures to single-phase BCC in refractory high-entropy alloy ZrHfNbTaW. XRD profiles in (a) overall view and (b) magnified view of (110) atomic planes for initial ingot ($N = 0$) and samples processed by HPT for $N = 1$, 10 and 50 turns. XRD profiles and Rietveld analysis for (c) initial ingot and (d) sample processed by HPT for $N = 50$ turns.

**Results**

XRD profiles are shown in Fig. 2(a) for the initial ingot and the samples processed by HPT for $N = 1$, 10 and 50 turns. The initial ingot contains two BCC phases with distinct peaks, but these peaks move closer to each other by HPT processing, as shown in a magnified view of XRD in Fig. 2(b). These peak shifts indicate that one BCC phase is expanded, and the other one is contracted. The lattice expansion of one phase and the contraction of another one by HPT processing, which was also reported in some other dual-phase HEAs such as AlFeCoNiCu [63], should be due to the gradual dissolution of the two phases in each other by straining. The material after $N = 50$ turns transforms into a single BCC phase. The detailed crystal structure analysis using Rietveld refinement, as shown in Fig. 2(c) for the initial ingot and in Fig. 2(d) for the sample processed by HPT for $N = 50$ confirms the occurrence of phase transformation by HPT processing. There are two BCC phases in the ingot: (i) BCC 1 with a lattice parameter of $a = 0.329$ nm and 85 wt% fraction, and (ii) BCC 2 with a lattice parameter of $a = 0.350$ nm and 15 wt% fraction. A single-phase structure is observed for the sample processed by HPT for $N = 50$ with lattice parameters of $a$ = 0.342 nm. This indicates the transformation of two BCC phases into a single-phase BCC structure with a lattice parameter between the lattice parameters of the initial two BCC phases. This transformation which needs a significant mass transfer is considered to be strain-induced and not pressure-induced [49,50,57,58].

Inverse pole figure (IPF) orientation mapping for the initial ingot obtained through EBSD analysis is shown in Fig. 3, in which the high-angle grain boundaries are shown with black color lines, and the low-angle grain boundaries are shown with white color lines. The grain size obtained through the line intercept method for the initial ingot of ZrHfNbTaW is 260 µm. Such large grain sizes are frequently observed in arc-melted HEAs [6,7,46-48].

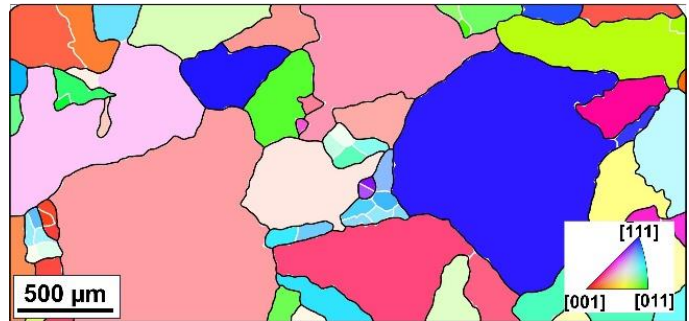

Figure 3. Coarse-grained structure with average grain size of 260 nm in refractory high-entropy alloy ZrHfNbTaW after arc melting. SEM-EBSD inverse pole figure orientation map in transverse direction for the initial ingot.

The SEM-EDS elemental mappings are shown in Fig. 4 for (a) the initial ingot and the samples processed by HPT for (b) 1, (c) 10 and (d) 50 turns. For the initial ingot, two apparent phases are revealed by SEM: a bright phase with 82 vol% and a dark phase with 18 vol%. The bright phase is rich in tantalum and tungsten, while the dark phase is rich in zirconium and hafnium. A comparison between the fraction of phases examined by XRD and SEM-EDS suggests that the bright and dark phases correspond to the BCC 1 and BCC 2 phases in Fig. 2(a), respectively. The two phases are deformed and mixed by HPT processing and the degree of mixing is enhanced by increasing the number of HPT turns from $N = 1$ to $N = 10$. After $N = 50$ turns, only a single phase is detected which is consistent with XRD analysis. The atomic composition determined by SEM-EDS is given in Table 1 for the initial ingot and for the samples processed by HPT for $N = 50$, indicating that despite differences in the composition of two initial BCC phases, the overall composition of both samples is consistent with the nominal composition. Elemental mixing (within the detection limit of SEM-EDS) and the transition of a dual-phase material to a single-phase alloy is a result of the applied ultra-SPD, as discussed in a few earlier publications [51,52].

To examine the mixing of elements by HPT at the nanometer level, STEM-EDS was conducted. The HAADF image and corresponding STEM-EDS mappings for ZrHfNbTaW after HPT processing for $N = 50$ is shown in Fig. 5. The STEM-EDS mappings confirm uniform mixing of all elements at the nanometer scale, and this is consistent with XRD and SEM analyses. The overall composition of the region selected in Fig. 5 is close to the nominal composition, indicating the uniform presence of all constituent elements after HPT processing for $N = 50$. Combining these STEM results with XRD analysis confirms the successful synthesis of a new single-phase alloy, adding a new alloy to the existing list of refractory HEAs [15-20].

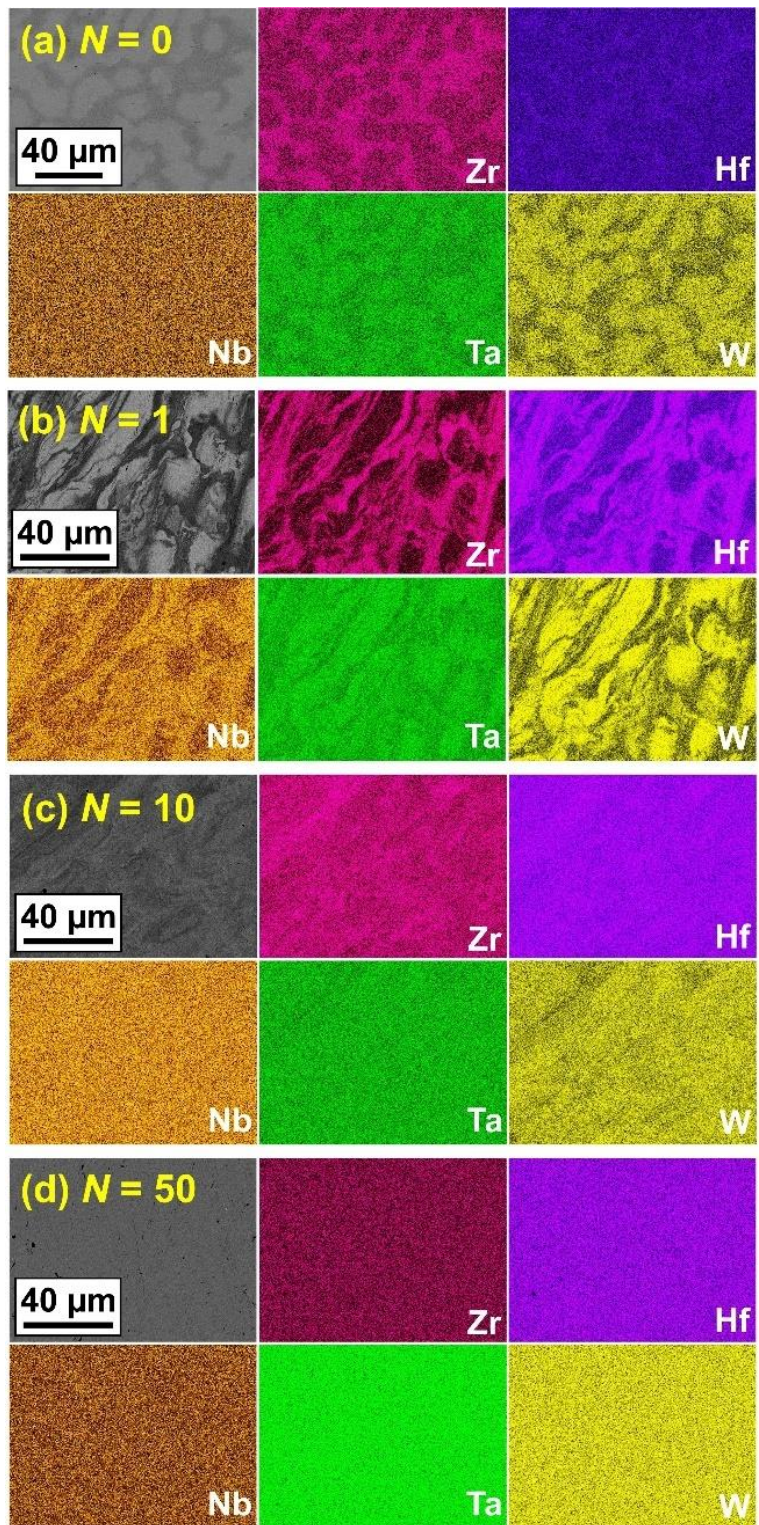

Figure 4. Strain-induced elemental mixing in refractory high-entropy alloy ZrHfNbTaW. SEM images and corresponding EDS elemental mappings for (a) initial ingot ($N = 0$) and samples processed by HPT for (a) $N = 1$, (b) $N = 10$ and (c) $N = 50$ turns.

Table 1. Strain-induced homogenization of composition in refractory high-entropy alloy ZrHfNbTaW. Atomic percentage of elements analyzed by SEM-EDS in different phases for initial ingot and sample processed by HPT for $N = 50$ turn.

|  | Measurement Method | Overall | Ingot BCC 1 Phase | BCC 2 Phase | HPT, $N = 50$ BCC 3 Phase |
|---|---|---|---|---|---|
| **Fraction (wt%)** | XRD |  | 85 | 15 | 100 |
| **Fraction (vol%)** | SEM |  | 82 | 18 | 100 |
| **Zr (at%)** | SEM-EDS | 18.8 | 10.2 | 31.1 | 21.3 |
| **Hf (at%)** | SEM-EDS | 22.3 | 19.2 | 27.5 | 21.1 |
| **Nb (at%)** | SEM-EDS | 20.7 | 22.8 | 16.6 | 20.6 |
| **Ta (at%)** | SEM-EDS | 19.5 | 24.9 | 12.1 | 18.4 |
| **W (at%)** | SEM-EDS | 18.8 | 22.9 | 12.8 | 19.6 |

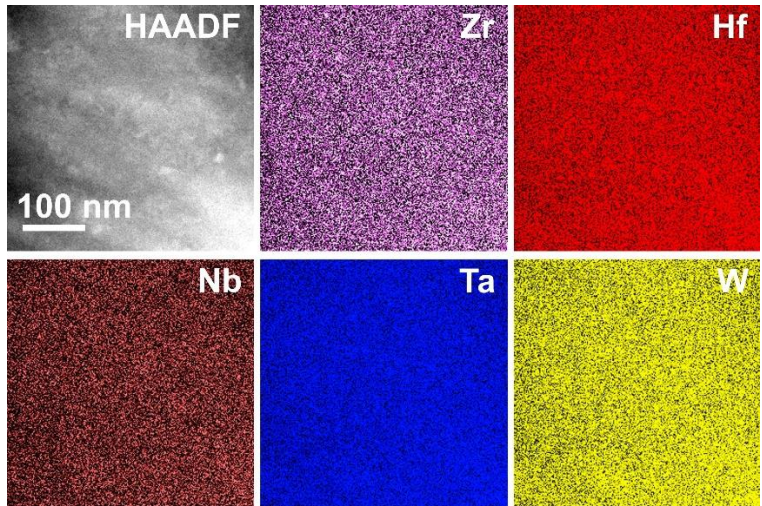

Figure 5. Uniform distribution of elements in refractory high-entropy alloy ZrHfNbTaW after ultra-SPD processing. HAADF micrograph and corresponding STEM-EDS elemental mappings for sample processed by HPT for $N = 50$ turns.

Microstructural analysis using TEM is shown in Fig. 6 for the sample treated by HPT for $N = 50$ turns, where (a) is a bright-field image, (b) is the corresponding selected area electron diffraction (SAED) analysis and (c) is the corresponding dark-field image. The SAED pattern confirms the presence of a single BCC phase which is consistent with XRD and SEM analysis. The SAED pattern is in a complete ring shape confirming the existence of nanograins with random misorientations in the chosen area. The presence of nanograins can be also confirmed in the bright-field image and more clearly in the dark-field image. The grain size distribution examined by the analysis of bright regions in several dark-field images is depicted in Fig. 6(d). The grains are at the nanometer scale, and the average grain size, taken from more than 100 grains, is around 12 nm which is much smaller than the grain sizes reported in conventional alloys processed using HPT [31,32,43,45,46,54,60,63] and is comparable to the ones reported in some HPT-processed ceramics [45].

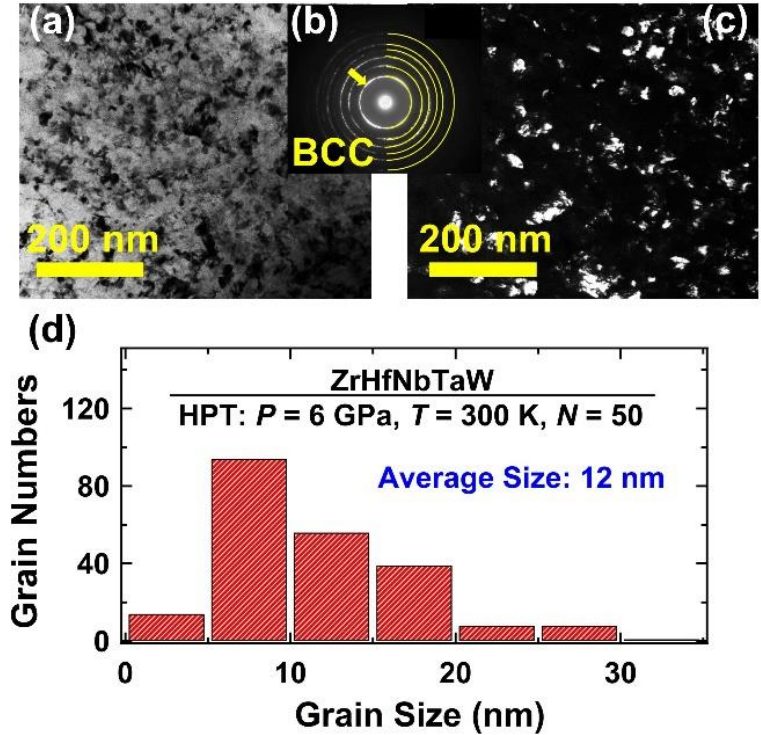

Figure 6. Formation of nanograins with average size of 12 nm in refractory high-entropy alloy ZrHfNbTaW after ultra-SPD processing. TEM (a) bright-field image, (b) SAED profile, (c) dark-field image, and (d) grain size distribution histogram for sample processed by HPT for $N = 50$ turns.

To examine the nanostructural features of ZrHfNbTaW after HPT processing for $N = 50$, high-resolution TEM images were taken as shown in Fig. 7 and Fig. 8. Examination of these figures indicates several important points. First, high-resolution TEM images and their FFT analysis confirm the presence of only the BCC phase, as indicated in Fig. 7(a) and 8(a). Second, despite small nanograin sizes, numerous dislocations are observed within some grains such as the ones shown in Fig. 7(a) using T marks or more clearly in Fig. 7(b) and 7(c) using lattice images. The estimated dislocation density for the four dislocations detected in Fig. 7(a) is $3.2 \times 10^{15}$ m$^{-2}$, which is comparable to highly deformed metals [36]. The presence of such a high dislocation density in a nanograined material is unusual as grain boundaries are expected to act as sinks for dislocations [64]. Dislocation survival within nanograins, which was also observed in some other HPT-processed HEAs [47] and ceramics [45], indicates the low mobility of dislocations in these materials. Third, the lattice is distorted even in regions with clear lattice fringes, such as the one shown in Fig. 8(b), possibly due to the presence of elements with different atomic sizes. Fourth, in addition to lattice distortion and dislocations, high-angle grain boundaries are detected in the microstructure as shown by dotted lines in Fig. 8(a) or more clearly in the lattice image of Fig. 8(c). These grain boundaries together with dislocations and lattice distortion are expected to provide a high hardness for this refractory HEA [34,35].

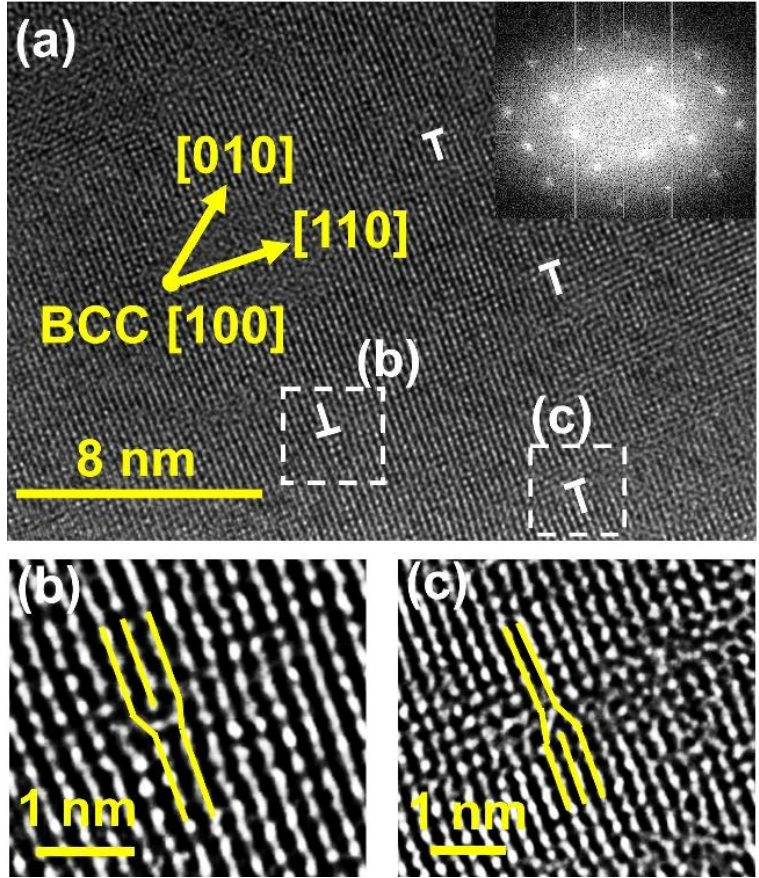

Figure 7. Formation of dislocations in refractory high-entropy alloy ZrHfNbTaW after ultra-SPD processing. TEM (a) high-resolution image and (b, c) lattice images for sample processed by HPT for $N = 50$ turns.

The microhardness of the initial ingot and samples processed for $N = 1$, 10 and 50 versus distance from the center of the discs is illustrated in Fig. 9(a). The microhardness of the initial sample is 560 Hv which is a high hardness for a coarse-grained material. The reason for such a high hardness should be due to the enhanced solution hardening mechanism in HEAs [13,14,17,18,22]. Fig. 9(a) shows that the hardness of HPT-processed discs increases as the distance from the center of the disc and the number of HPT turns increase. These variations of hardness can be justified by shear strain changes because shear strain increases with increasing both the distance from the disc center and the number of HPT turns, as mentioned in Eq. (1) [27,44]. The variation of hardness versus shear strain for the four samples is summarized in Fig. 9(b). The hardness increases with increasing shear strain, but changes in hardness versus strain at large strains become less abrupt compared to changes at low strains. The reason for less significant hardening at large strains is due to the contribution of dynamic recovery [53,54], dynamic recrystallization [55,56], and grain boundary migration [55,56]. The maximum hardness of 860 Hv is achieved at 4 mm away from the center of the disc processed by HPT for $N = 50$. This hardness value is 1.5 times more than the hardness of the initial ingot and higher than the hardness levels reported in other BCC-type refractory HEAs [15–19].

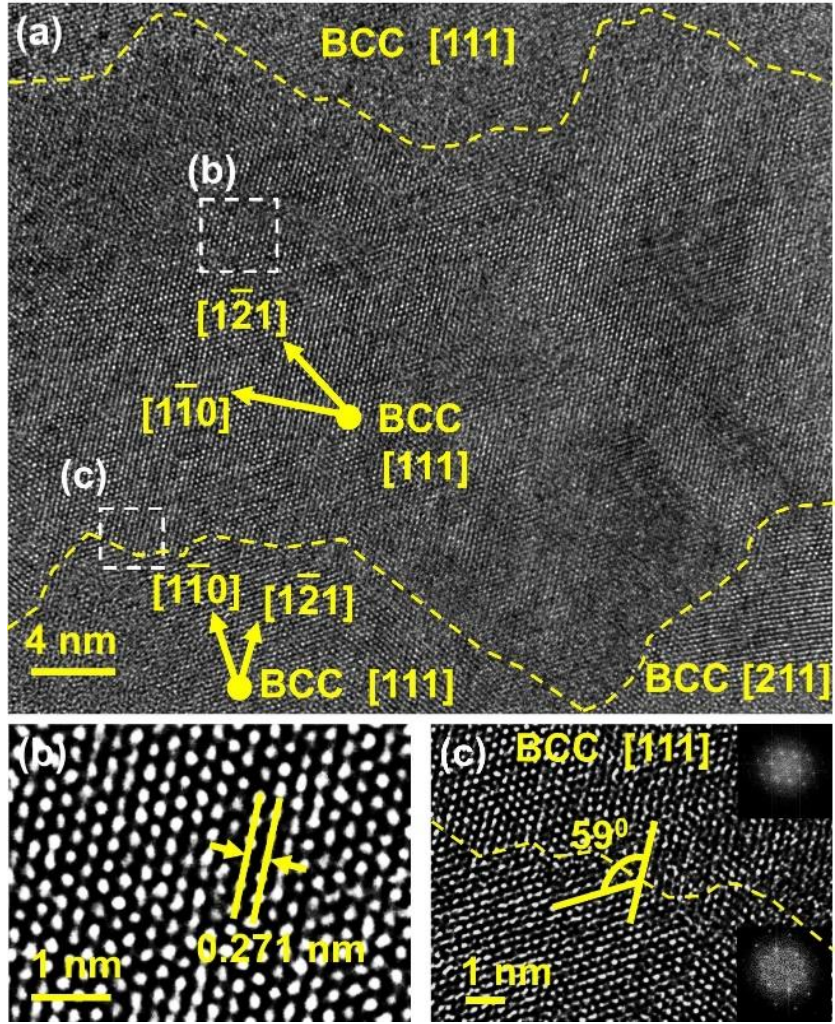

Figure 8. Formation of distorted lattice and high-angle grain boundaries in refractory high-entropy alloy ZrHfNbTaW after ultra-SPD processing. TEM (a) high-resolution image and (b, c) lattice images for sample processed by HPT for $N = 50$ turns.

The average melting temperature of the elements in ZrHfNbTaW is 2874 K and the melting temperature predictor software estimates a melting temperature of 2908 K [61]. This melting temperature is higher than the values reported for other refractory HEAs in the literature [13–19,21–24]. DSC analysis conducted to up to 873 K (the maximum temperature achievable in the authors' facility) confirmed the absence of any exothermic or endothermic peaks, and post-DSC analysis by XRD confirmed the high phase stability in this high-melting-point HEAs. The high thermal stability of this alloy should be due to the sluggish diffusion effect reported in HEAs, while this effect is expected to be more significant in this alloy because of its high melting point [1-5]. The theoretical density for single-phase ZrHfNbTaW alloy after HPT processing for $N = 50$ is 12.1 g/cm$^3$ by considering its crystal structure and composition, a value which is close to the experimental density of 12.3 g/cm$^3$ measured using the Archimedes method for the ingot.

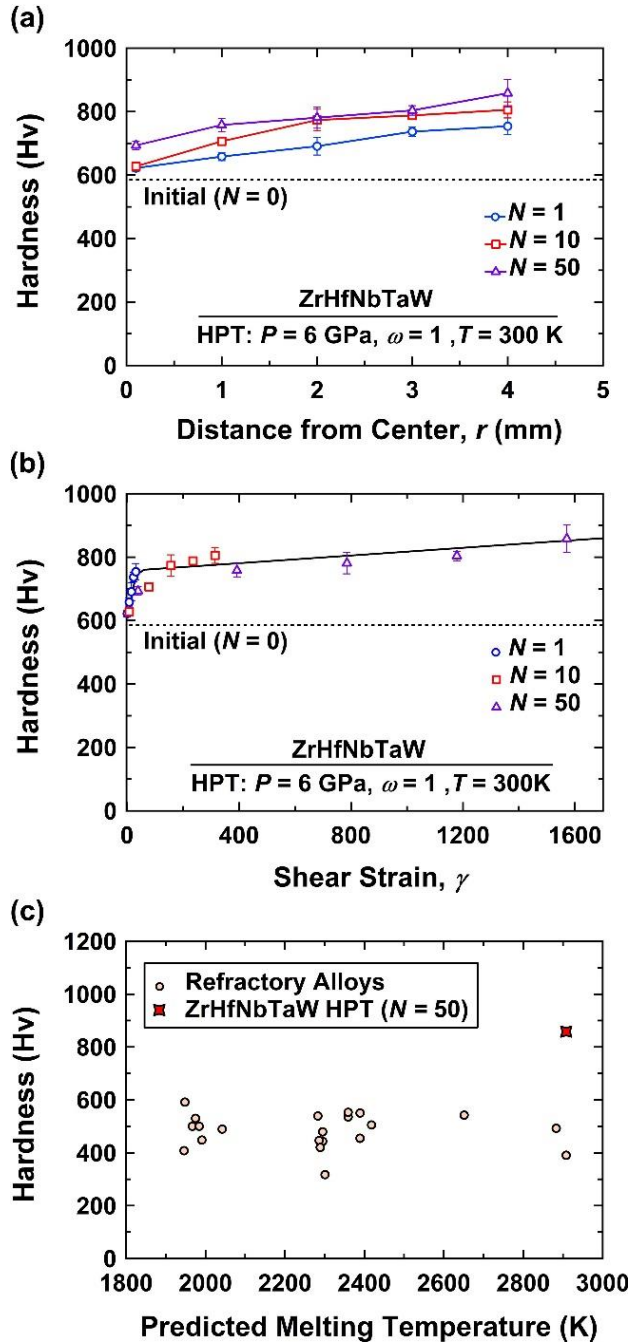

Figure 9. Ultrahigh hardness and high melting point of high-entropy alloy ZrHfNbTaW after ultra-SPD processing. Hardness variations versus (a) distance from disc center and (b) shear strain for initial ingot ($N = 0$) and samples processed by HPT for $N = 1$, 10 and 50 turns. Hardness variations versus melting point for the sample processed by HPT for $N = 50$ turns in comparison with refractory alloys reported in literature.

**Discussion**

This study introduces a new alloy ZrHfNbTaW with a high melting point, ultrahigh hardness and nanostructure which can have potential applications as a refractory material. Three issues in this research must be discussed further: (i) the performance of ZrHfNbTaW among the existing refractory HEAs, (ii) the reason behind the ultrahigh hardness, and (iii) the mechanism of the formation of a single-phase BCC structure.

In response to the first issue, it can be mentioned that the present ZrHfNbTaW alloy after HPT treatment exhibits markedly better hardness in comparison to other refractory HEAs and MEAs such as W-Nb-Mo-Ta [13], W-Nb-Mo-Ta [13], HfMoTaTiZr [14], HfMoNbTaTiZr [14], $Nb_{25}Mo_{25}Ta_{25}W_{25}$ [15], $V_{20}Nb_{20}Mo_{20}Ta_{20}W_{20}$ [15], MoNbTaTiV [16], $Ta_{20}Nb_{20}Hf_{20}Zr_{20}Ti_{20}$ [17], NbTaTiV [18], NbTaVW [18], NbTaTiVW [18], MoNbTaTi [19], MoNbTaTiW [19], $NbCrMo_{0.5}Ta_{0.5}TiZr$ [22], $AlMo_{0.5}NbTa_{0.5}TiZr$ [23], Al-Nb-Ta-Ti-Zr [24], Al-Hf-Nb-Ta-Ti-Zr [24] and Al-Nb-Ta-Ti-V-Zr [24]. A comparison between the hardness versus predicted melting temperature of other refractory HEAs is given in Fig. 9(c). The ZrHfNbTaW alloy after HPT processing for $N = 50$ lies on the upper end for both hardness and melting point. This indicates the remarkable potential of this HEA for refractory applications. Processing this alloy by HPT is essential in this study because it not only leads to ultrahigh hardness but also to a single-phase BCC structure formation. It is known from the literature that among the main metallic crystal structures, BCC has the highest potential for high-temperature application [21–24]. The predicted yield strength for ZrHfNbTaW after HPT processing for $N = 50$ is 2.8 MPa by considering its hardness and the Tabor relationship, and this is higher than the strength of conventional refractory metals such as vanadium, niobium, molybdenum and tungsten [12]. The yield stress of 2.8 MPa is also over two times higher than the yield stress of refractory HEAs reported in the literature [13-19,22]. It should be noted that some HPT-processed HEAs exhibit a hardness higher than the one achieved for this alloy, but those HEAs have dual phases and usually contain an ordered phase [46-48]. Among bulk single-phase HEAs, the current alloy exhibits an exceptional hardness level. Despite the promising hardness and melting point of ZrHfNbTaW and its good thermal stability examined by DSC, its mechanical properties under high temperatures need to be investigated in future works as the final target of refractory materials is the application under high temperatures [14,16,17].

Concerning the second question about the mechanism underlying ultrahigh hardness, three main mechanisms can contribute to the hardening of this alloy. This HEA is made up of five transition metals with different atomic sizes and high elastic modulus. This mixing gives an intrinsic severe lattice distortion and enhanced solution strengthening to the HEA which leads to a high microstructural resistance to dislocation motion [2]. Lattice distortion and trapping of dislocations in the microstructure are observable in TEM high-resolution images in Figs. 7 and 8. This enhanced solution hardening is responsible for the high hardness of the initial coarse-grained ingot whose hardness is about three times higher than the prediction by the rule of mixture. The same observation is reported in other HEAs in which solution hardening is more significant than levels expected from conventional Fleischer and Labusch models [13,14,17,18,22]. After HPT processing, additional strengthening is achieved by grain-boundary hardening and dislocation hardening [65]. Grain-boundary hardening, which is described by the Hall-Petch mechanism [34,35], should be significant in this HEA because the average grain size after HPT processing for $N = 50$ turns reaches 12 nm. Generally, hardening by dislocation is negligible in nanograined materials because grain boundaries act as a sink for dislocations, and thus, dislocations cannot survive within nanograins [64]. However, a large density of dislocations is observed in HPT-processed nanograined ZrHfNbTaW, as shown using high-resolution TEM in Fig. 7. The presence of these dislocations, which should also contribute to significant hardening, should be due to the resistance of the distorted structure of HEAs to dislocation motion as well as due to sluggish diffusion as a core effect of HEAs [1-5]. Although it is not possible to quantify the effect of each strengthening mechanism on the overall hardness of this HEA due to a lack of data about its physical parameters, it can be mentioned that solution hardening, grain-boundary hardening and dislocation hardening contribute to the high hardness.

Regarding the third question, this study shows that a refractory HEA with a single BCC phase is successfully synthesized by HPT processing for $N = 50$ turns, while the initial ingot has a dual-phase crystal structure containing two BCC phases. Analyses by XRD, SEM and STEM suggest that the initial BCC phases are dissolved in each by shear straining resulting in the formation of a new BCC phase with a lattice parameter ranging between the lattice parameters of the two initial phases. Such phase transformation which has a mechanical alloying nature is controlled by shear strain during HPT processing [66]. It was shown that when shear strain is so large (usually, $\gamma > 1000$), many materials can transform into single-phase structures even in some immiscible systems [49,50]. Under such large strain conditions, known as ultra-SPD [49,50], the high applied strain and the continuous formation of lattice defects change the dynamic stability of phases and result in the formation of phases that cannot be synthesized under normal processing conditions [57,58]. Although these phase transformations are strain-induced, high pressure in HPT is also essential for three reasons. First, high pressure allows plastically deform such hard materials which their processing is not possible under normal conditions [45]. Second, high pressure suppresses dynamic atomic diffusion resulting in retaining the new phases during the process [67]. Third, theoretical studies based on multiscale atomistic and continuum theories showed that stress tensor concentration on lattice defects, such as dislocations, occurs by straining under high pressure, leading to fast phase transformation and generation of new phases [57,68]. In summary, HPT appears to be an effective method for developing new single-phase and intermetallic-free refractory HEAs, opening new research directions to investigate the high-temperature properties of these materials for different applications. Future studies should focus on the large-scale production of such refractory materials for practical applications, utilizing the various upscaled SPD methods currently available [36, 42].

**Conclusions**

In this study, a novel refractory high-entropy alloy (HEA) ZrHfNbTaW was synthesized by arc melting flowed by high-pressure torsion (HPT) processing. The initial HEA had dual body-centered cubic (BCC) phases, but it transformed into a single-phase BCC structure after HPT processing for 50 turns. The alloy synthesized by HPT exhibited an ultrahigh hardness of 860 Hv due to the combined effect of severe lattice distortion and solution hardening, grain-boundary hardening, and dislocation hardening. This study introduces the HPT method as an effective process to explore new refractory HEAs.

**Acknowledgment**

The author S.D. thanks the MEXT, Japan for a scholarship. This work is supported in part by grants-in-aid for scientific research from the MEXT, Japan (JP22K18737).